\documentclass{aa}
\usepackage{graphicx}
\usepackage{txfonts}
\usepackage{amssymb}
\usepackage[]{amsmath}
\usepackage{nccmath}
\usepackage{natbib}
\usepackage{booktabs}
\usepackage{mathtools}
\usepackage{physics}
\usepackage{multirow} 
\usepackage[export]{adjustbox}
\usepackage{hyperref}
\usepackage{subfig}
\usepackage{wrapfig}
\usepackage{float}
\usepackage{graphicx}
\usepackage{gensymb}
\hypersetup{colorlinks,linkcolor={black},citecolor={blue},urlcolor={black}} 
\def\degree{^{\circ}}
\def\expval#1{\langle#1\rangle}

\begin{document}

   \title{X-ray analysis of the accreting supermassive black hole in the radio galaxy \object{PKS 2251+11}}

   \author{S. Ronchini
          \inst{1,2}
          \and 
          F. Tombesi\inst{2,3,4,5}
          \and
          F. Vagnetti\inst{2}
          \and
          F. Panessa\inst{6}
          \and
          G. Bruni\inst{6}
          }

   \institute{Gran Sasso Science Institute (GSSI), Viale Francesco Crispi, 7 - 67100 L'Aquila (AQ), Italia
         \and
             Department of Physics, University of Rome “Tor Vergata”, Via della Ricerca Scientifica 1, I-00133 Rome, Italy
        \and
            Department of Astronomy, University of Maryland, College Park, MD 20742, USA 
        \and
            NASA/Goddard Space Flight Center, Code 662, Greenbelt, MD 20771, USA
        \and 
            INAF-Osservatorio Astronomico di Roma, Via Frascati 33, 00078 Monteporzio Catone, Italy
        \and
            INAF-Istituto di Astrofisica e Planetologia Spaziali, via Fosso del Cavaliere 100, I-00133 Roma, Italy
             }
  \date{Received 31 January 2019 / Accepted 13 March 2019}
 
  \abstract
   {We have investigated the dichotomy between jetted and non-jetted active galactic nuclei (AGNs), focusing on the fundamental differences of these two classes in the accretion physics onto the central supermassive black hole (SMBH). We tested the validity of the unification model of AGNs through the characterization of the mutual interaction between accreting and outflowing matter in radio galaxies.}
   {Our aim is to study and constrain  the structure, kinematics and physical state of the nuclear environment in the broad line radio galaxy (BLRG) \object{PKS 2251+11}. The high X-ray luminosity and the relative proximity make such AGN an ideal candidate for a detailed analysis of the accretion regions in radio galaxies. The investigation will help to shed light on the analogies and differences between the BLRGs and the larger class of radio-quiet Seyfert galaxies and hence on the processes that trigger the launch of a relativistic jet.}
   {We performed a spectral and timing analysis of a $\sim$64 ks observation of PKS 2251+11 in the X-ray band with XMM-Newton. We modeled the spectrum considering an absorbed power law superimposed to a reflection component. We performed a time-resolved spectral analysis to search for variability of the X-ray flux and of the individual spectral components.}
   {We find that the power law has a photon index $\Gamma=1.8\pm 0.1$, absorbed by an ionized partial covering medium with a column density $N_H=(10.1\pm 0.8) \times 10^{23}$ cm$^{-2}$, a ionization parameter $\log{\xi}=1.3\pm 0.1$ erg s$^{-1}$ cm and a covering factor $f\simeq90\%$. Considering a density of the absorber typical of the broad line region (BLR), its distance from the central SMBH is of the order of $r\sim 0.1$ pc. An Fe K$\alpha$ emission line is found at 6.4 keV, whose intensity shows variability on timescales of hours. We derive that the reflecting material is located at a distance $r\gtrsim600r_s$, where $r_s$ is the Schwarzschild radius. }
   {Concerning the X-ray properties, we found that \object{PKS 2251+11} does not differ significantly from the non-jetted AGNs, confirming the validity of the unified model in describing the inner regions around the central SMBH, but the lack of information regarding the state of the very innermost disk and SMBH spin still leaves unconstrained the origin of the jet.}


   \maketitle

%


\section{Introduction}
Accretion onto supermassive black holes (SMBHs) is the physical mechanism powering active galactic nuclei (AGNs). As such, these sources are very efficient at converting the gravitational energy of the accreting mass in other forms of energy, as radiation or powerful outflows. The primary radiation emitted by the accreting material is reprocessed by the surrounding nuclear environment, so that the AGN is observable at all wavelengths, from radio to $\gamma$-rays. While there is consensus about what powers AGN, the structure of the nuclear environment is still a matter of debate. The unified model of AGNs (e.g. \citealp{1993ARA&A..31..473A}) posits that all active galaxies share a common structure, which can be summarized with a few fundamental components: i) a central SMBH; ii) an accretion disk;  iii) a dusty obscuring torus, typically on the same plane of the accretion disk; and iv) outflows, in the form of highly collimated relativistic jets, or wide-angle nuclear winds (\citealp{2017A&ARv..25....2P}). According to the unified model, the distinction between unobscured (Type 1) and obscured (Type 2) AGNs is related to the inclination of the line of sight with respect to the plane of the accretion disk (\citealp{1995PASP..107..803U}). Radio-loud quasars and less-luminous radio galaxies are capable of launching powerful, relativistic jets, while radio-quiet QSOs and Seyfert galaxies have jets that are much less powerful (the role of jet production in the AGN classification is reviewed by \citealp{2017NatAs...1E.194P}). The mechanism responsible for the launching and the collimation of the relativistic jet is an open issue, and is likely related to the interaction between the magnetic field and the accretion disk (e.g. \citealp{2017SSRv..207....5R}, \citealp{2010ASPC..429...91K}).\par  
In this regard, X-ray observations provide a powerful tool for the investigation of the accretion-ejection physics in the neighborhood of the SMBH (e.g. \citealp{2007ASPC..373..125M}, \citealp{2009A&ARv..17...47T}, \citealp{2016AN....337..404R}). While non-jetted AGNs are relatively well studied in the X-rays, we have a limited sample of radio galaxies whose X-ray properties are well constrained. For this reason we cannot yet confidently affirm that the underlying structure of the nuclear environment is the same for both non-jetted and jetted AGNs (e.g. \citealp{2007ApJ...664...88G}). Since the fraction of absorbed AGN is comparable among non-radio galaxies and radio galaxies (\citealp{2016MNRAS.461.3153P}), they could share a common obscuring structure, as predicted by the unified model. The presence or not of systematic differences in the X-ray properties between these two populations would help to clarify the role of the accretion processes in the triggering of the jet. AGNs show similarities with stellar BH binaries (\citealp{2003MNRAS.345.1057M}), where the X-ray luminosity and the hardness of the X-ray spectrum are related with the production of jets and winds (\citealp{2012MNRAS.422L..11P}, \citealp{2016LNP...905...65F}).\par
In this regard, the broad line radio galaxies (BLRGs) are especially suitable for high quality X-ray studies, given that they have the highest X-ray fluxes in the radio-loud population. The inclination angle between the line of sight (l.o.s.) and the perpendicular to the accretion disk in BLRGs has an intermediate value, meaning that the radiation coming out from the nuclear region is neither suppressed by the obscuring torus, nor overwhelmed by the relativistically boosted emission of the jet, as in blazars. Thanks to this characteristic, we are able to observe directly the optical emission of the broad line region (BLR) and we can also investigate the structure, the geometry, and the physics of the nuclear environment down to sub-pc scales.\par
Since the early observations with X-ray telescopes, radio loud (RL) AGNs appear slightly more luminous in the X-ray band, compared to the radio quiet (RQ) counterpart with comparable optical luminosity (\citealp{1987ApJ...313..596W}, \citealp{2018A&A...619A..95C}). RL AGNs have also slightly flatter power law continuum, with $\expval{\Gamma_{RL}}\sim \expval{\Gamma_{RQ}}-0.5$ (e.g. \citealt{1997MNRAS.288..920L}). However, these early results were affected by the low sensitivity of the instruments. For instance, \cite{1999ApJ...526...60S} found only a weak indication that the 2-10 keV spectra of BLRGs are flatter than those of Seyferts 1 galaxies. \cite{1998MNRAS.299..449W} found in a sample of five BLRGs an average Fe K$\alpha$ equivalent width (EW) of $\sim 100$ eV and a Compton hump significantly weaker compared to those observed in RQ AGNs. The weakness of reflection components in BLRGs was confirmed by \cite{2000ApJ...537..654E}, who proposed that this effect may be due to a small solid angle subtended by the disk to the primary continuum. This scenario is compatible with two-temperature disks, optically thick in the outer part and truncated in the inner part. The truncation may be induced by an advection-dominated accretion flow (ADAF) or ion-supported tori.\par
On the other hand, \cite{2002MNRAS.332L..45B} suggest that the weakness of reflection features is not necessarily due to geometrical factors. Instead, they propose that the reprocessing by an ionized accretion disk could produce a similar effect, discarding the hypothesis that the RL-RQ dichotomy, and hence the production of the jet, is due to an intrinsic difference in the disk accretion state. In fact, other fundamental parameters could play a significant role in the production of the jet, such as high accretion rates and/or high values of mass and spin of the SMBH. Nonetheless, it is still not clear which combination of these parameters is the most favorable to trigger the jet emission (\citealp{Chia}, \citealp{Marc}). \cite{2007MPLA...22.2397B} claims that the accretion geometry of RL AGNs could be related to the accretion efficiency, in turn linked to the AGN luminosity. In particular, for bolometric luminosities $L_{bol}<0.01 L_{Edd}$ the accretion transits to ADAFs, explaining the absence of reflection by the inner part of the disk ($L_{Edd}$ is the Eddington luminosity). For $L_{bol}>0.01 L_{Edd}$, the disk is not truncated and the reflection can extend down to the innermost stable circular orbit (ISCO). The presence of reflecting material close to the SMBH is responsible for the emission of a relativistically broadened Fe K$\alpha$ line, whose evidence is mixed in the case of BLRGs. The Fe K$\alpha$ line can be originated by several components, such as the accretion disk, the BLR, or the inner part of the obscuring torus. In BLRGs each component can contribute differently, producing a final Fe K$\alpha$ line width that varies case by case (e.g. \citealp{2010ApJ...719..700T}, \citealp{ogl}, \citealp{2013ApJ...772...83L}, \citealp{2009ApJ...700.1473S}, \citealp{2015ApJ...814...24L}).\par
In the discussion of \cite{2009ApJ...700.1473S}, the authors infer that there is also a dichotomy in the outflow typology between BLRGs and Seyfert galaxies. In this scenario, accretion disk winds would be produced only by Seyferts, while the jet is a prerogative of radio galaxies. If we suppose that the bulk of accretion energy is used for winds acceleration, the jet formation is prevented, thus explaining the lack of jets in Seyferts. Nevertheless, the subsequent work of \cite{Tomb_2014} controverts this scenario, demonstrating that the production of mildly-relativistic disk winds is not a prerogative of RQ AGNs. In fact ultra fast outflows (UFOs) are found in both RQ and BLRGs (\citealt{Tomb_2010}, \citealt{Tomb_2014}, \citealt{goff}, \citealt{ree_go}, \citealp{2015ApJ...813L..39L}), indicating that their origin is somewhat independent of the formation of the relativistic jet. On the other hand, the presence of this latter remains a prerogative of RL galaxies.\par
Recent X-ray observations with XMM-Newton, Suzaku and NuSTAR (\citealp{2011ApJ...734..105S}, \citealp{2014ApJ...794...62B}, \citealp{2015ApJ...808..154R}, \citealp{2018MNRAS.478.2663U}) suggest that the accretion structure in BLRGs and in Seyfert galaxies is similar, therefore indicating that the separation in different classes is subtle. Instead, on the basis of these last results, the first ones can be considered as a sub-class of the second ones.

\section{PKS 2251+11}
\label{ar_data}
\object{PKS 2251+11} is a BLRG with a redshift $z=0.3255$, derived from optical emission lines. \cite{2009ApJ...705..298M}, modeling the infrared spectrum emitted by the dusty torus, derived the angle $i$ between the direction of the jet and the l.o.s., obtaining $i=67\degree$. This result indicates that \object{PKS 2251+11} may be at the border between Type 1s and Type 2s, possibly intercepting part of the obscuring torus. Since the authors do not indicate the uncertainty associated with the estimation of $i$, in our analysis we will assume an inclination angle contained in the range $50\degree<i<70\degree$. Besides, the authors estimated an opening angle of the torus equal to $\sigma=15\degree$, whose value is one of the smallest among the other sources in the sample. \par
\cite{1991A&A...241....5W} used two independent methods for estimating the mass of the central SMBH $M_{BH}$ in this object, obtaining $M_{BH}\simeq1.4\times 10^9 M_{\odot}$ and $M_{BH}\simeq1.7\times 10^9 M_{\odot}$, with a fractional difference of $\sim 17\%$. In the same article the authors also derived $\log{[\eta \dot{M}c^2/L_{Edd}}]=-1.52$ and $\dot{M} \simeq 1 M_{\odot}/\text{yr}$ ($\dot{M}$ is the mass accretion rate and $\eta$ is the accretion efficiency).\par
\cite{2011ApJ...728...98D}, using a dynamical approach, estimated $M_{BH}\simeq7.2 \times 10^8 M_{\odot}$. Assuming the optical emission dominated by the accretion disk, the authors derive also a mass accretion rate of $\dot{M}=4.6 \, M_{\odot}yr^{-1}$, which is not far from the value obtained by \cite{1991A&A...241....5W}. Other derived quantities are $L_{bol}=(1.35 \pm 0.03)\times 10^{46}$ erg/s and an accretion efficiency $\eta=0.066$. \cite{2002ApJ...571..733V}, through single-epoch spectroscopic measurements, obtained $M_{BH}=(9.1^{+3.5}_{-2.5})\times 10^8 M_{\odot}$, comparable both with the estimates given by \cite{2011ApJ...728...98D} and \cite{1991A&A...241....5W}. In the following we will consider only the value given by \cite{2002ApJ...571..733V}, as it is likely the most accurate estimate.\par
\cite{2008MNRAS.390..847L}, studying the correlation between the radio and X-ray luminosities in a sample of both RL and RQ AGNs, report for \object{PKS 2251+11} these quantities: $L_R=\nu L_{\nu}|_{\nu=\text{6 cm}}=1.15 \times 10^{43}$ erg/s, where $L_R$ is the total radio luminosity at 6 cm (taken from \citealp{1989AJ.....98.1195K}, which also found that the contribution of the core to the radio luminosity is less than 1$\%$);
$L_X=1.15 \times 10^{44}$ erg/s, with $L_X$ is the X-ray luminosity in the energy range E=0.2-20 keV; $\alpha_{ox}=0.372 \log{[F_{\nu=\text{2 keV}}/F_{\nu=\text{3000\AA}}]}=-1.86$; $\log{R}=2.56$, where $R$ is the radio loudness, equivalent to $R=1.36 \times 10^5 L_5/L_B$, with $L_5=L_{\nu}|_{\nu=\text{5 GHz}}$ and $L_B=L_{\nu}|_{\nu=\text{4400\AA}}$.
\cite{1999ApJ...526...60S}, performing an X-ray spectral analysis of RL AGNs observed with ASCA, reported  radio archival data of \object{PKS 2251+11}. They derived the parameter $\log{{\mathcal{R}}}=[\text{radio core power}/\text{lobes radio power}]=-1.51$, which indicates that the core contributes only to 3$\%$ of the total radio luminosity. In the same article the X-ray spectrum of \object{PKS 2251+11} is analyzed and the adopted model is an absorbed power law. The derived best fit parameters are $\Gamma=1.14^{+0.23}_{-0.18}$ and $N_H=6.0^{+3.0}_{-2.2}\times 10^{21}\text{cm}^{-2}$. Especially for $E<2$ keV, absorption and emission features appear rather complex. The authors tried to add a Gaussian line in correspondence of the Fe K$\alpha$ line, keeping fixed during the fit the energy at 6.4 keV and the width at $\sigma_{ga}=50$ eV. They derive an upper limit of the equivalent width of EW$<315$ eV.

\section{Data reduction}

\object{PKS 2251+11} was observed with XMM-Newton on 2011 December 18, for a total exposure of 63.8 ks. In this work we consider the EPIC-pn data. The detector operated in full window imaging mode. Data reduction and analysis were performed following the SAS standard procedure\footnote{described by the threads on the web page \url{https://www.cosmos.esa.int/web/xmm-newton/sas-threads}}. In order to check the presence of particle flaring background, we considered the entire field of view and we selected only the photons with energy in the range E=10-12 keV. Then we plotted a light curve of the selected photons, excluding from the initial observation all the intervals of time in which the count rate is higher than 0.6. Subsequently, we extracted a circular region centered on the source, with a radius of 35''. The background region is of the same shape and dimension but centered in an area free of emission from point sources, located in the same CCD. The test for the presence of pile-up gave a negative response. After the filtering procedure, the net exposure of the observation was reduced to 57.3 ks. Along with the cleaned spectrum, a redistribution matrix file (RMF) and an ancillary response function (ARF) were created with standard SAS tools. The spectral data were grouped choosing a minimal threshold of 25 photons per bin and a $\chi^2$ statistic is adopted. Only in the case of time resolved spectroscopy (see below) the minimal threshold is lowered to five or ten photons per bin and a C-statistic is adopted.  Unless otherwise specified, the errors are reported with at $1\sigma$ level of confidence, while the upper and lower limits are reported with a 90$\%$ level of confidence.

\section{Spectral analysis}
Our analysis is performed using the Heasarc v. 6.21 and XSPEC v. 12.9.1.
We began trying to fit the data with an absorbed power law, following the standard assumption that a hot corona comptonizes the soft seed photons emitted by the accretion disk (e.g. \citealt{Haa}).
\subsection{Neutral total covering absorber}
First we included only the absorption due to the interstellar medium (ISM) in our Galaxy, for which \cite{dickey} estimate a column density $N_H^{\text{Gal}}=0.052\times 10^{22}\text{cm}^{-2}$. For the estimate of the absorption cross section, we refer to the Tuebingen-Boulder model (\citealp{tbabs}), which in XSPEC is indicated as \texttt{tbabs}. The fit gives a $\chi^2/d.o.f.=917.5/369$. Therefore the Galactic absorption is not sufficient to fit the spectrum and we added an absorption component due to neutral gas in the AGN host galaxy, but this does not improve the fit ($\chi^2/d.o.f.=917.5/368$).
\subsection{Neutral partial covering absorber}
\begin{figure}
        \centering
        \includegraphics[width=1\columnwidth]{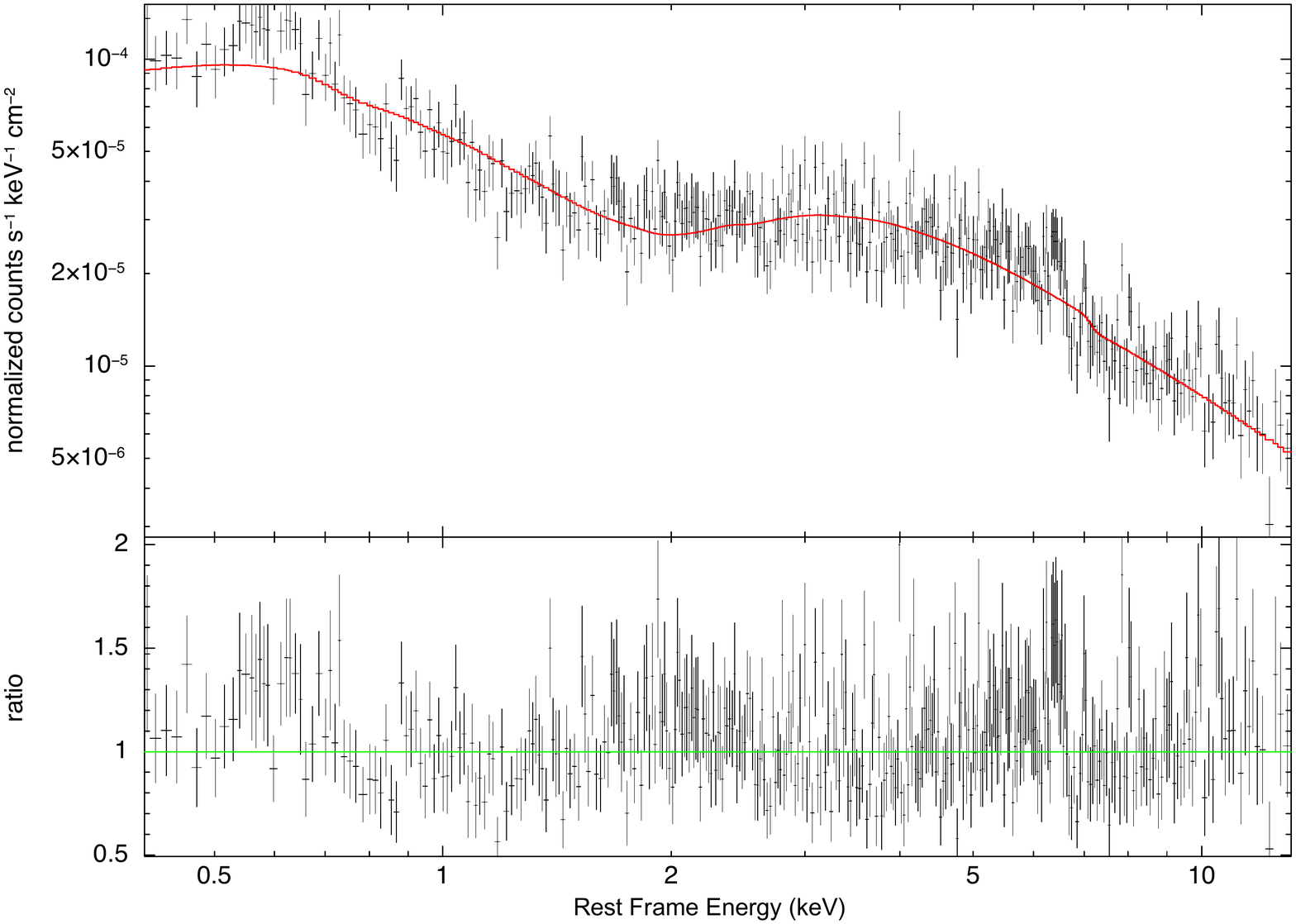}
        \centering
        \caption{Comparison between the EPIC-pn spectrum and the model \texttt{zpcfabs*tbabs*zpo}. The data to model ratio is showed in the bottom panel.}
        \label{mo2}
\end{figure}

Then we tested the hypothesis of a neutral partial covering absorption, modeled in XSPEC as \texttt{pcfabs}. The resulting model is written in XSPEC language as \texttt{zpcfabs*tbabs*zpo}. Adding just one parameter, the fit improves to a value $\chi^2/d.o.f.=448.3/367$. The validity of this improvement is also confirmed by the ftest, which gives a confidence larger than 5$\sigma$. The derived best fit parameters are $N_H=7.08      _{-  0.30   }^{+  0.31  } \times 10^{22}\text{cm}^{-2}$, $f=0.85\pm 0.01$ and $\Gamma=1.76     \pm 0.04$, where $N_H$ is the column density of the gas in the source host galaxy, $f$ is the covering factor, in other words, the fraction of the solid angle subtended by the absorbing medium, and $\Gamma$ the photon index of the power law. The comparison between this model and the EPIC-pn spectrum is showed in Fig. \ref{mo2}.\par
\begin{figure}
        \centering
        \includegraphics[width=1\columnwidth]{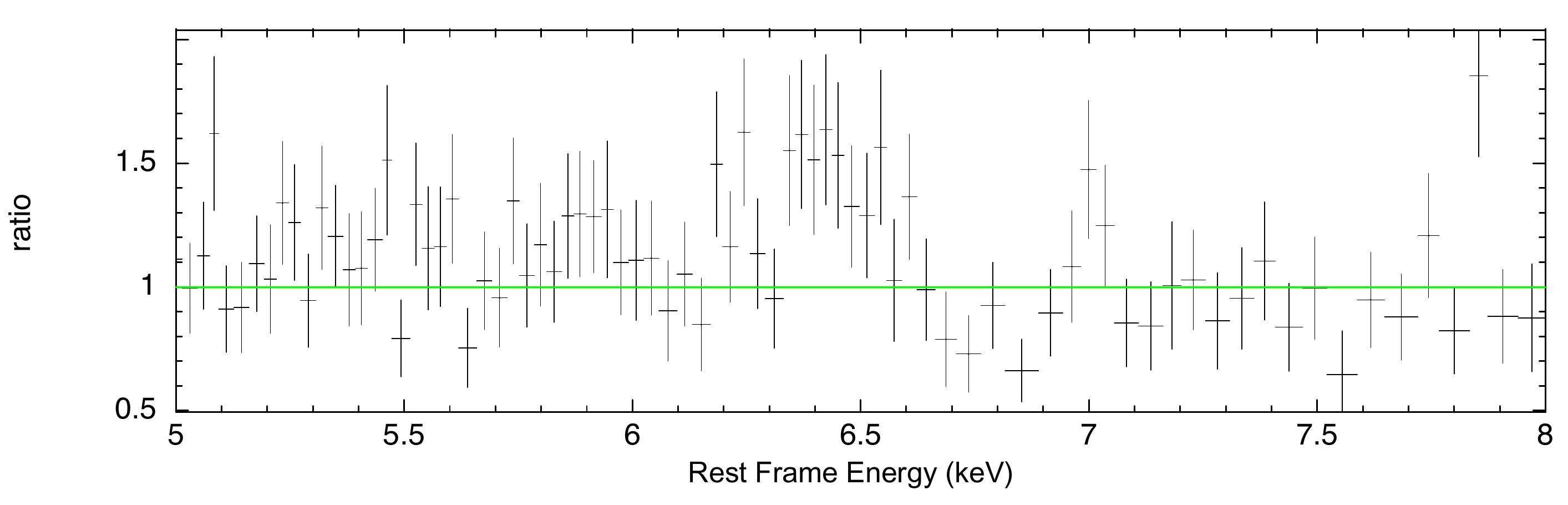}
        \centering
        \caption{Zoom of the data to model ratio in the Fe K band at energies E=6-7 keV for zpcfabs model. The bump around 6.4 keV suggests a possible neutral Fe K$\alpha$ emission line.}
        \label{zoom_ga}
\end{figure}

\begin{figure}
        \centering
        \includegraphics[width=1\columnwidth]{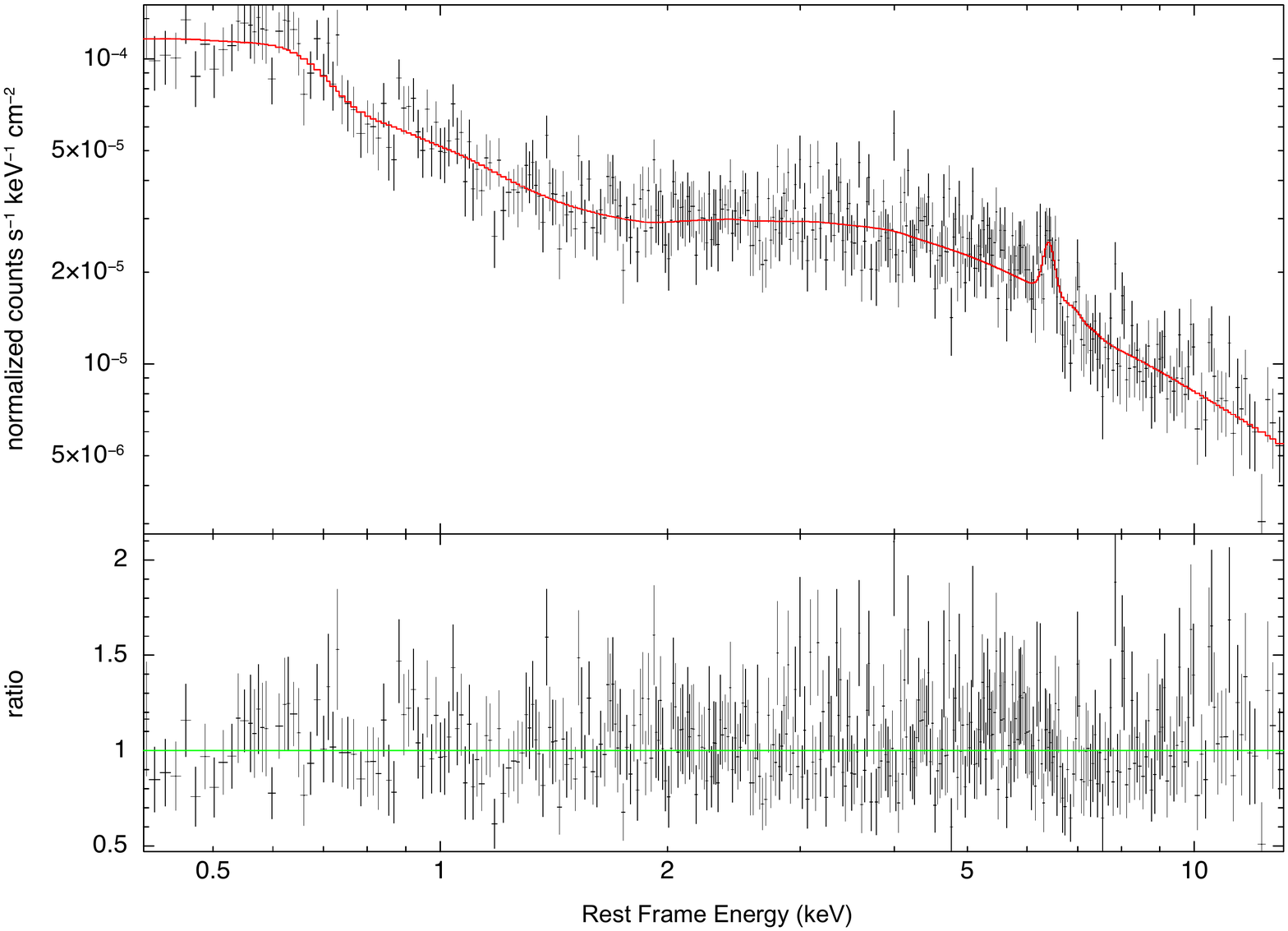}
        \centering
        \caption{Comparison between the EPIC-pn spectrum and the model \texttt{zxipcf*tbabs*(zpo+zga)}. With respect to the Fig. \ref{mo2} the agreement improves especially in the band E=0.5-1 keV, where the presence of ionization lines and edges most influences the overall shape of the spectrum. The data to model ratio is showed in the bottom panel.}
        \label{4_1}
\end{figure}

\begin{table}
\centering
\footnotesize
\caption{Values of the best-fit model including a Gaussian emission line for the Fe K$\alpha$ or pexmon. $N_{\text{PL}}$ indicates the normalization of the power law at 1 keV, $N_{ga}$ is the normalization of the Gaussian line and $z_{\text{PL}},z_{abs},z_{ga}$ are the redshift of the power law continuum, the absorber and the Gaussian line, respectively. The parameters denoted with a (*) are fixed during the fit.}
\renewcommand{\arraystretch}{1.5}
\begin{tabular}{l r r c}

\hline\hline
Parameter & Gaussian  & pexmon & Units\\\hline

      $N_H^{Gal}$*             & $ 0.052$  &$ 0.052$&$\times 10^{22}\text{cm}^{-2}$     \\
      $N_H$             & $10.1 \pm 0.8$ & $9.0     \pm  1.0 $ &$\times 10^{22}\text{cm}^{-2}$      \\
      $\log{\xi}$            & $1.33_{-  0.06  }^{+  0.08   }$ & $1.31 \pm 0.06 $&$ \,\text{erg}\,\text{s}^{-1}\,\text{cm}$\\
      f                 & $0.88      \pm 0.01$ &$0.87 _{-  0.02 }^{+  0.01   }    $ &- \\
      $\Gamma$             & $1.78     \pm 0.05$&$1.79      \pm 0.05$&-       \\
      $N_{\text{PL}}$     & $7.3      _{-  0.8  }^{+  0.9   }$& $4.1_{-  0.4  }^{+  0.5   }$& $\times10^{-4}\frac{\text{ph}}{\text{keV} \, \text{cm}^{2}\text{s}} $     \\
      $E_{ga}$     &  $6.41      \pm 0.03$& - &keV                \\
      $\sigma_{ga} $*            & $10$  & - &eV    \\
      $N_{ga}$  & $3.6      \pm 0.8$& - & $\times10^{-6}\frac{\text{ph}}{\text{cm}^{2}\text{s}} $  \\
      $R$ & - & $1.03^{+0.35}_{-0.30}$ &-\\
      i* &- & 60$\degree$ &- \\
      $z_{\text{PL}}=z_{abs}=z_{ga}$*   & $ 0.33$  &$ 0.33$&-     \\
\hline\hline

\end{tabular}
\label{xi_pex}
\end{table}

\subsection{The Fe K$\alpha$ line}
As can be seen in Fig. \ref{zoom_ga}, there is an excess of counts between 6 and 7 keV, which we identify with the Fe K$\alpha$ fluorescence line. This last is modeled with a Gaussian profile, with a centroid $E_{\text{ga}}$ and a width $\sigma_{ga}$. The model is written as \texttt{zpcfabs*tbabs*(zpo+zga)}. Letting both the parameters $E_{\text{ga}}$ and $\sigma_{ga}$ vary freely during the fit, we find that the line is unresolved, since we obtain only an upper limit for $\sigma_{ga}\leq 120$ eV. The inclusion of this component improves the fit to  $\chi^2/d.o.f.=424.8/364$. The best fit value for the energy is $E_{\text{ga}}=6.41\pm 0.03$ keV. The significance for the addition of this line is $>$3.7 $ \sigma$. In the following, since the line is unresolved, we fixed the width $ \sigma_{ga}$ at 10 eV.
\subsection{Ionized partial covering absorber}
A further improvement of the fit occurs if we relax the hypothesis of neutral absorbing gas, which is no more valid if the primary continuum is particularly intense and/or if the absorber is located close to the X-ray source. In this case we substituted in XSPEC the model \texttt{pcfabs} with the model \texttt{zxipcf} (\citealp{Ree}), which introduces the ionization parameter $\xi=L_{\text{ion}}/(nr^2)$, where $L_{\text{ion}}$ is ionizing luminosity in the energy range E=13.6 eV - 13.6 keV, while $n$ and $r$ are the density and the distance form the SMHB of the absorber, respectively. The fit gives a high improvement to $\chi^2/d.o.f.=382.9/364$. The best fit values are $\log{\xi}=1.33_{-  0.06  }^{+  0.08   }\text{erg}\,\text{s}^{-1}\,\text{cm}$, $N_H=10.1     _{- 0.81   }^{+  0.83 }\times 10^{22}\text{cm}^{-2}$, $f=0.88      \pm 0.01$ and $\Gamma=1.78      \pm 0.05$. The comparison between this model and the spectrum is showed in Fig. \ref{4_1}. Until now we have assumed that the redshift $z$ of the source and that of the absorber $z_{\text{abs}}$ are the same, namely that this last is at rest in the source reference frame. Then, if we let free $z_{\text{abs}}$ during the fit, we can explore the possibility that the absorber has a motion along the line of sight in the rest frame of the source. However, given the low energy resolution of the EPIC-pn and relatively low soft X-ray flux of the source, we are not able to statistically constrain the kinematics of the absorber.\par
From the final best-fit model including zxipcf we derive an absorbed flux in the energy range E=0.5-10 keV equal to $F^{\text{abs}}_{0.5-10\, keV}=(1.33^{+0.02}_{-0.03})\times 10^{-12} \text{erg }\text{cm}^{-2}\text{s}^{-1}
$. The unabsorbed flux is $F^{\text{unabs}}_{0.5-10\, keV}=(2.5^{+0.1}_{-0.2})\times 10^{-12} \text{erg }\text{cm}^{-2}\text{s}^{-1}$, which is the flux emitted by the primary continuum without considering the the suppression due to absorption. The corresponding intrinsic X-ray luminosity in the energy range E=0.5-10 keV is $L^{\text{int}}_{0.5-10\, keV}=(8.1\pm 0.5)\times 10^{44} \text{  erg }\text{s}^{-1}$. Therefore the absorption causes a significant decrease of the intrinsic flux equal to $[F^{\text{unabs}}-F^{\text{abs}}]/ F^{\text{unabs}} \simeq 47 \, \%$.
\subsection{The pexmon model}
Finally we tried to investigate the nature of the Fe K$\alpha$ emission line using the reflection model \texttt{pexmon} (\citealt{2007MNRAS.382..194N}) which introduces the parameter $R=\Omega/2\pi$, namely the fraction of the solid angle covered by the reflecting medium, seen by the primary X-ray source. We assume a solar abundance, a high-energy cutoff at 150 keV, an inclination angle $i=60\degree$ (in the middle of our confidence range $50\degree<i<70\degree$), while we let free $\Gamma$, $R$ and the parameters of the absorber. We derive $\log{\xi}=1.31_{-  0.06  }^{+  0.07   }\text{erg}\,\text{s}^{-1}\,\text{cm}$, $N_H=9.03     _{- 0.90   }^{+  0.98 }\times 10^{22}\text{cm}^{-2}$, $f=0.87      \pm 0.01$,  $\Gamma=1.79      \pm 0.05$ and $R=1.03_{- 0.30   }^{+  0.35 }$. Even assuming an inclination angle in the range between 50$\degree$ and 70$\degree$, the reflection factor does not change significantly, being in any case compatible with 1. The results from zxipcf and pexmon models are summarized in Table \ref{xi_pex}. \par

\section{Timing analysis}

\subsection{X-ray variability}

In order to check for X-ray variability, we created a light curve of the count rate in the energy band E=0.3-10 keV. For different bin times (100 s, 500 s, 1000 s, 5000 s and 10000 s) we have calculated: i) the average value $\expval{C}$ of the count rate; ii) the $\chi^2$ probability $p$, which indicates the probability that the observed variability is due to chance. As a threshold, we consider significant the variability only when $1-p>99.7\%$ ($3\sigma$ level of confidence); iii) the $\sigma_{var}=[C_{max}-C_{min}]/\sqrt{\sigma_{max}^2+\sigma_{min}^2}$, with $C_{max}$ and $C_{min}$ the highest and lowest points in the light curve, while $\sigma_{max}$ and $\sigma_{min}$ the respective errors of $C_{max}$ and $C_{min}$; iv) the normalized excess variance $
\sigma_{NXS}^2=[S^2-\expval{\sigma^2}]/\expval{C}^2$, where $S^2$ is the variance of the light curve, that is $S^2=[\sum_i C_i^2-\expval{C}^2]/N$, where the sum is extended over all the $N$ points and $\expval{C}$ is the average value of the count rate; $\expval{\sigma^2}$ is the average value of the errors associated to each point.
\begin{table}
\centering
\begin{tabular}{ l c c c c }
\centering
bin time& $\expval{C}$ & $p$ & $\sigma_{var}$ & $\sigma_{NXS}^2$ \\
sec      & counts/s     & -        & -           & $\times 10^{-5}$\\
\hline
100      & 0.2461       & $82\%$   & 4.0         & -3         \\
500      & 0.2461       & $41\%$   & 3.7         & -99         \\
1000     & 0.2467       & $41\%$   & 3.8         & -72         \\
5000     & 0.2472       & $46\%$   & 1.6         & -15         \\
10000    & 0.2484       & $53\%$   & 1.5         & +7        \\
\hline
\end{tabular}
\caption{Variability estimator values for different choices of bin time.}
\label{tab_lc}
\end{table}

\begin{figure}
  \centering
  \includegraphics[width=1\columnwidth]{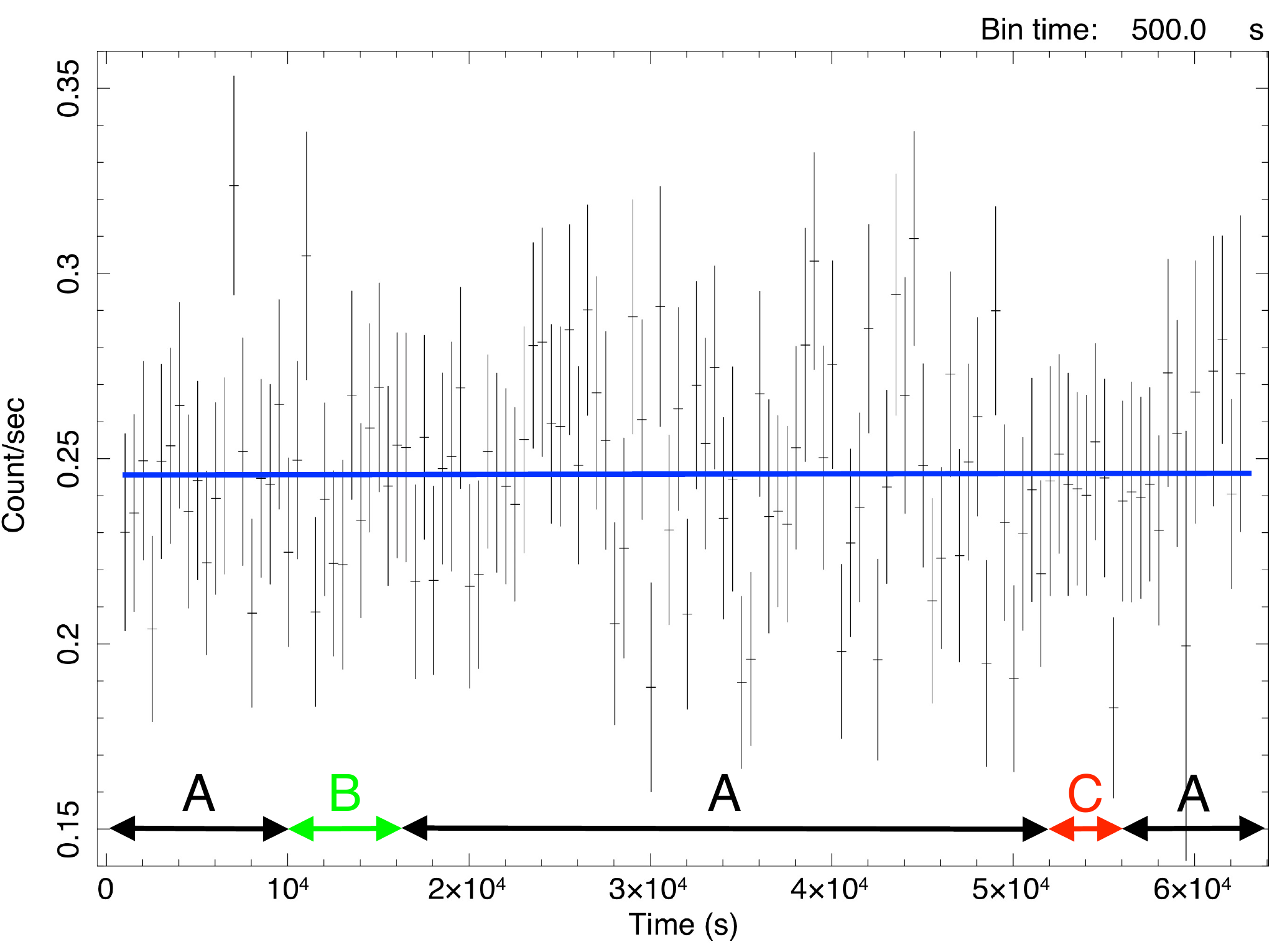}
  \centering
  \caption{Light curve of the photon count rate in the energy range E=0.3-10 keV, from the EPIC-pn observation. The horizontal blue line indicates the average value of the count rate. A, B, and C indicate the time intervals defined in Section \ref{trsa}.}
  \label{lc_500}
\end{figure}
For a bin time of 500 s we report the light curve of the count rate in Fig. \ref{lc_500}. From Table \ref{tab_lc} we can see that the variability is not significant for any choice of bin time.

\subsection{Time resolved spectral analysis}
\label{trsa}
\begin{figure}
        \centering
        \includegraphics[width=1\columnwidth]{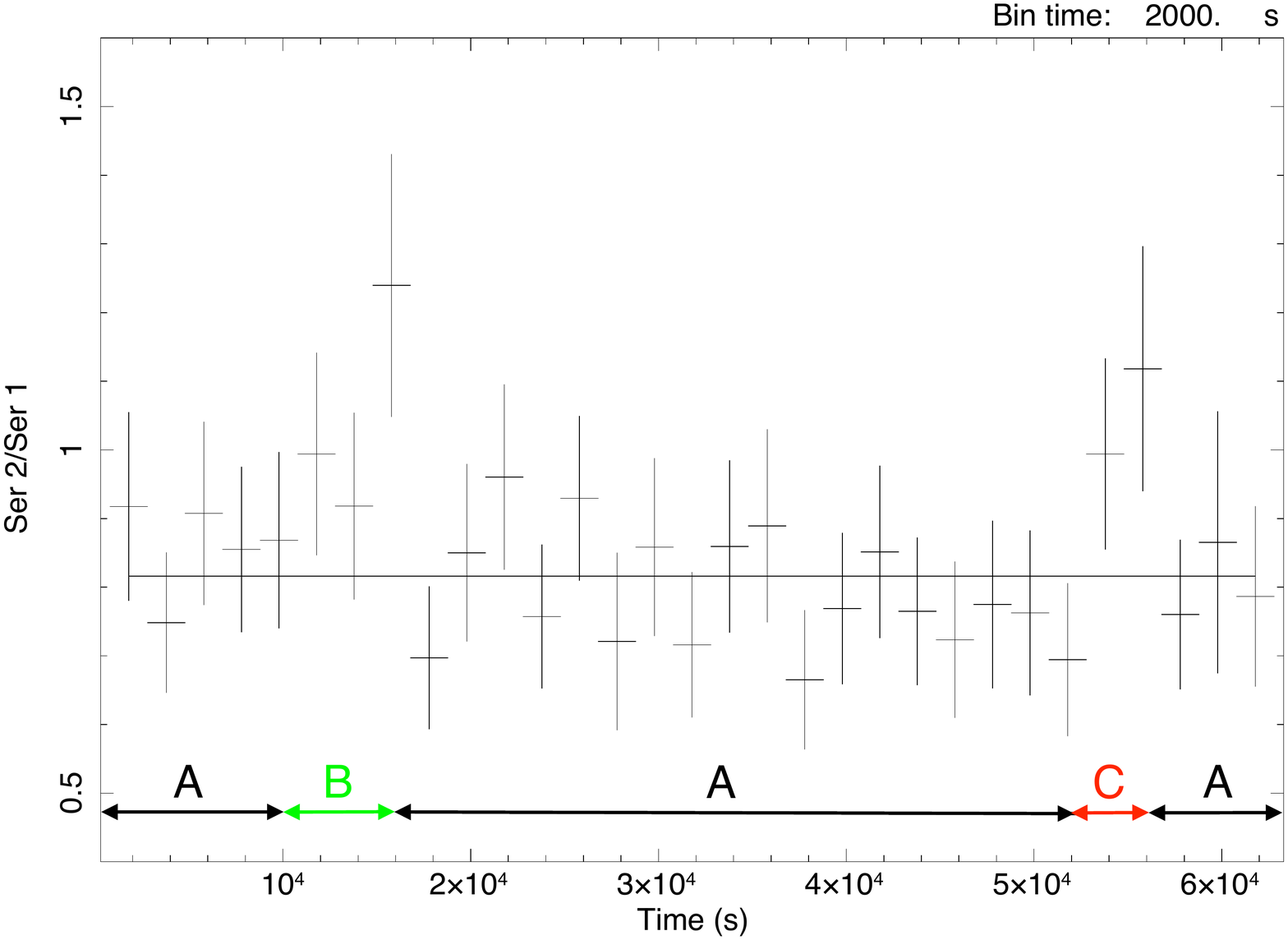}
        \centering
        \caption{Hardness ratio $HR$ as a function of time, for a bin time choice of 2000 s. The horizontal solid line is the average value of $HR$. Around 16 ks and 55 ks the $HR$ differentiates from its average value. A, B, and C indicate the time intervals defined in Section \ref{trsa}.}
        \label{HR}
\end{figure}

At this point, we tested the variability of the individual spectral components through a time resolved spectral analysis. First, we have produced two light curves, one in the soft X-ray band E=0.5-2 keV and the other in the hard X-ray band E=5-10 keV (the energies are referred to the source reference frame). Then we computed the hardness ratio as $
HR(t)=C_{5-10\text{ keV}}(t)/C_{0.5-2\text{ keV}}(t)
$, where $C_{5-10\text{ keV}}(t)$ and $C_{0.5-2\text{ keV}}(t)$ are the hard and soft light curves, respectively. From the Fig. \ref{HR}, the $HR$ is almost constant apart in some time intervals where it increases during a time scale of $10^4$ s. We divided the whole exposure in three sub-intervals, namely the interval A $\rightarrow$ $[0\text{ ks}<t<10\text{ ks}]$ + $[16 \text{ ks}<t<52\text{ ks}]$ + $[56\text{ ks}<t<62\text{ ks}]$; the interval B $\rightarrow$ $10\text{ ks}<t<16\text{ ks}$; the interval C $\rightarrow$ $52\text{ ks}<t<56\text{ ks}$.
For each of these intervals we have extracted the spectrum. Since the intervals B and C have a very short exposure, the collected photons are $\sim$1/10 of those of the total observation, with subsequent poor statistic for spectral parameters determination. So we first tried to merge B and C, gaining in S/N, but loosing information on the individual intervals. The new interval BC has [$10\text{ ks}<t<16\text{ ks}$] + [$52\text{ ks}<t<56\text{ ks}$]. We analyzed at the same time the spectra from A and BC, with the model \texttt{zxipcf*tbabs*(zpo+zga)}. Apart from $N_H^{Gal}$, $\sigma_{ga}$ and the redshift of the components, all the other parameters are left free during the fit. No parameter changes significantly from A to BC, apart for the normalization of the Fe K$\alpha$ line, which increases in BC, when the hardness ratio is higher.\par
Then we analyzed A, B, and C separately. Since the counts in B and C are few, if we let the spectral parameters for A, B, and C vary freely, the fit does not converge, or the errors are large. So we assumed that, apart the normalization of the Fe K$\alpha$, all the parameters do not change during the observation, obtaining
$ N_{ga}^A = 2.8^{+1.5}_{-1.4} \times10^{-6}\frac{\text{ph}}{\text{cm}^{2}\text{s}}, N_{ga}^B = 7.9^{+5.1}_{-4.4}\times10^{-6}\frac{\text{ph}}{\text{cm}^{2}\text{s}}, N_{ga}^C <9.2 \times10^{-6}\frac{\text{ph}}{\text{cm}^{2}\text{s}}$, where the superscript refers to the line normalization in A, B and C, respectively.

\begin{figure}
  \centering
    
  \includegraphics[width=.9\columnwidth]{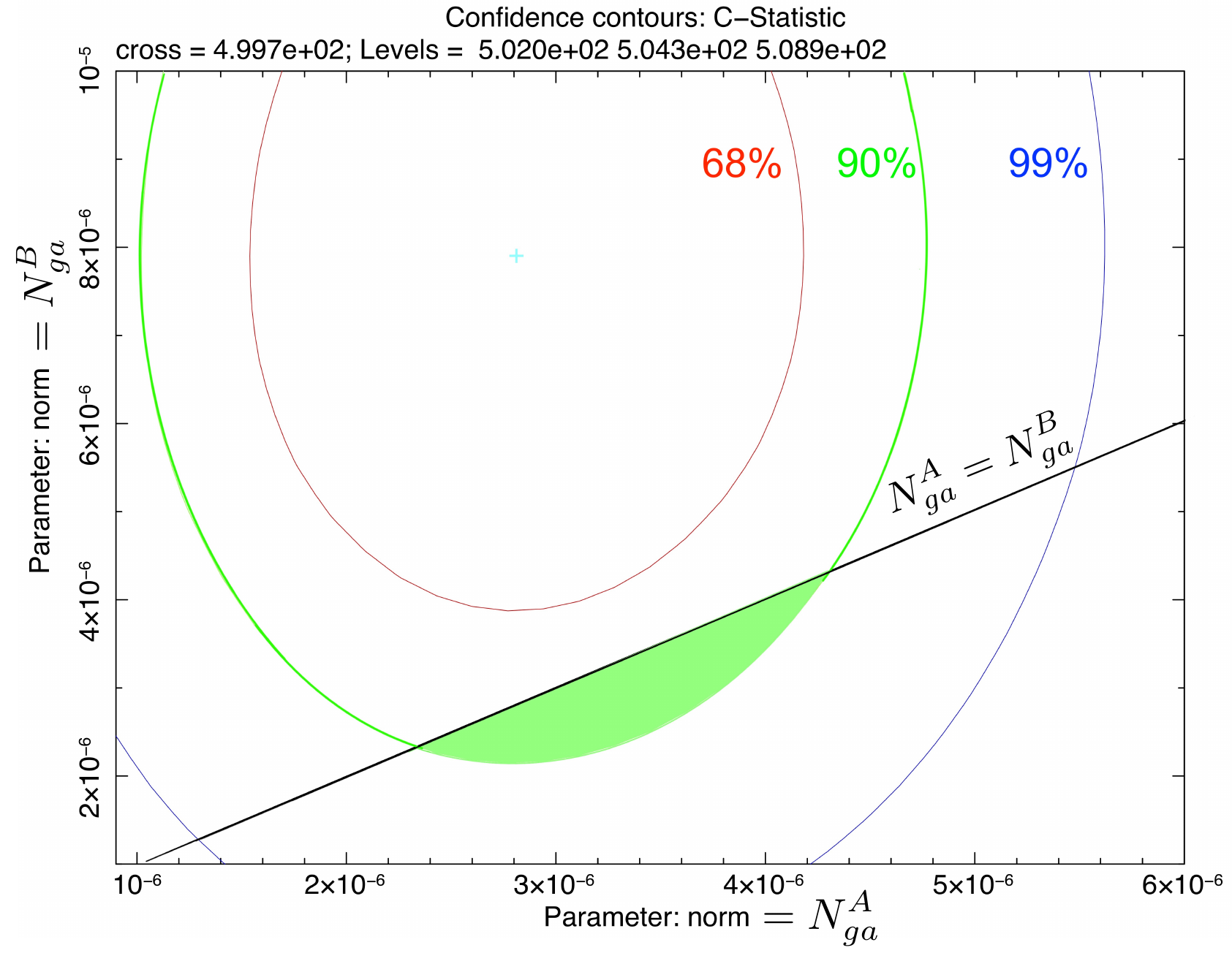}
  \caption{Plot of the confidence contours for $N_{ga}^A$ and $N_{ga}^B$. The straight black line is drawn in correspondence of $N_{ga}^B=N_{ga}^A$. The green area is the intersection between the half-plane below the black line and the region contained inside the $90\%$ contour.}
  \label{cont_norm}
\end{figure}

Apparently, within the errors, in A, B, and C the $N_{ga}$ is compatible with a constant. Although, since the exposure of C is too short, we cannot have a precise constraint on $N_{ga}$, while the variability from A to B is more remarkable. In order to test the variability of $N_{ga}$ between A and B, we considered the contour plot of $N_{ga}^A$ and $N_{ga}^B$ 
in Fig. \ref{cont_norm}, where we drew the line $N_{ga}^A=N_{ga}^B$. The green region shows the area in the parameter space where $N_{ga}^A>N_{ga}^B$ with a 90$\%$ level of confidence. As evident, it is considerably more probable that $N_{ga}^B>N_{ga}^A$. As a further check, we tried to use the model \texttt{zxipcf*tbabs*pexmon}, following the same procedure described above, fixing all the parameters equal for A, B and C and letting free only the reflection parameter $R$. In A, B and C the reflection parameter assumes the values $R_A=0.74^{+0.49}_{-0.38}$, $R_B=1.61^{+0.87}_{-0.75}$ and $R_C=0.79^{+0.89}_{-0.45}$, respectively. They are all compatible within the errors, but looking at the contour plot of $R_A$ and $R_B$ we notice that the intersection between the 90$\%$ area and the region where $R_A>R_B$ is extremely small. Therefore we can affirm that also the reflection parameter varies sensibly from A to B. Such result is in agreement with the increasing of $N_{ga}$ in the interval B. In fact, the greater is the solid angle covered by the reflector, the higher is the intensity of the reflection component. Finally we tried to vary the inclination angle $i$ between 50$\degree$ and 70$\degree$, always finding that $R_B\sim 2 \times R_A$, while $R_C \sim R_A$.

\begin{figure}
  \centering
    
  \includegraphics[width=.9\columnwidth]{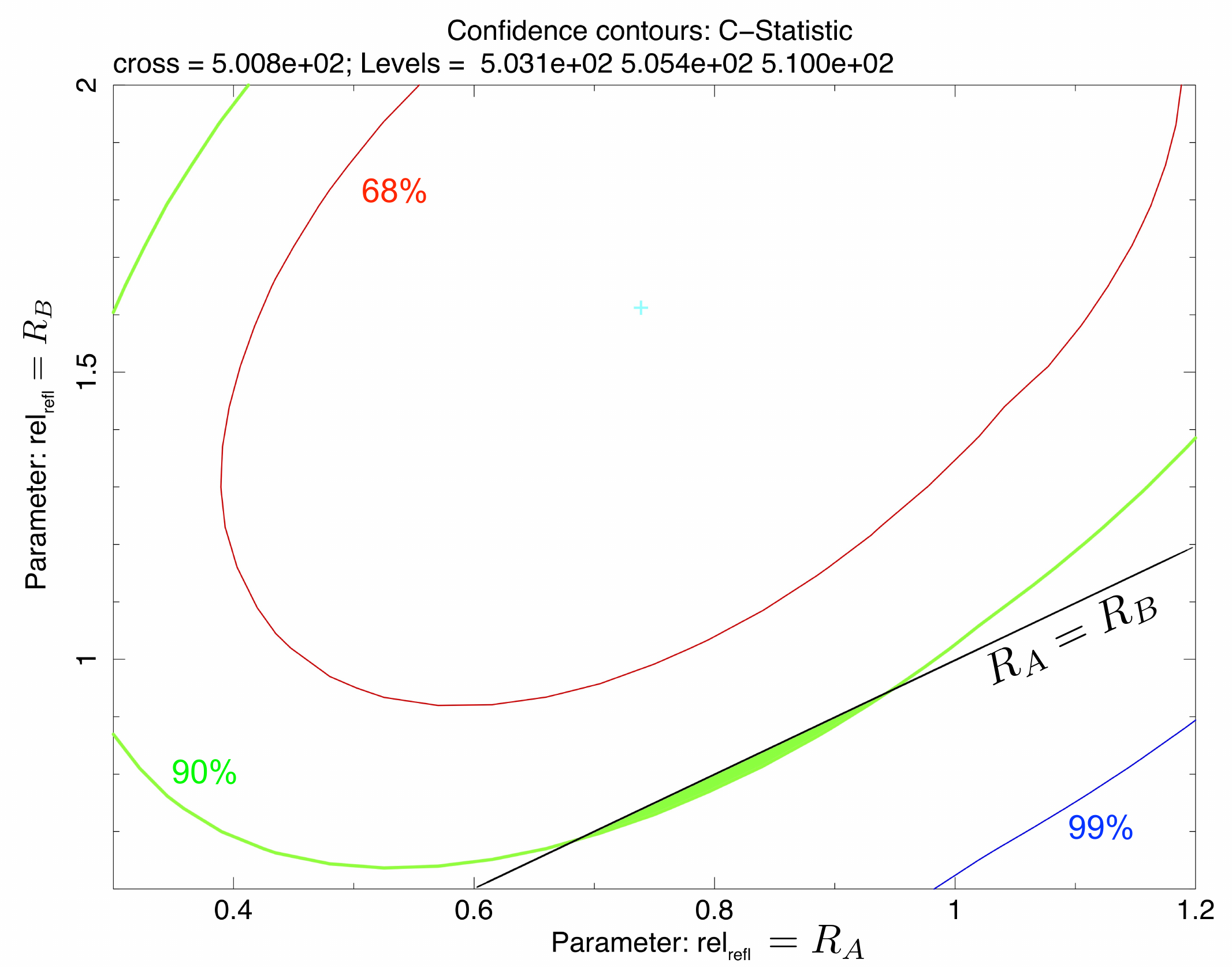}
  \caption{Plot of the confidence contours for $R_A$ and $R_B$. The straight black line is drawn in correspondence of $R_A=R_B$. The green area is the intersection between the half-plane below the black line and the region contained inside the $90\%$ contour.}
  \label{cont_pex}
\end{figure}

\section{Results and discussion}
In our spectral analysis of the X-ray spectrum of \object{PKS 2251+11}, the best-fit is obtained considering an absorbed power law, where the absorbing medium is ionized and partially covers the primary X-ray source. We also detect the presence of an unresolved Fe K$\alpha$ emission line at 6.4 keV.
\subsection{Location of the absorber}
The ionization parameter $\xi=\frac{L_{\text{ion}}}{nr^2}$, used in the photoionization models, contains information about the physical state of the absorber, which is likely to be stratified along the line of sight. The stratification implies that the gas has a value of $\xi$ that changes with $r$. In this case $r$ and $n$ are defined as the inner radius of the absorber and the corresponding density, respectively. Usually the ionizing luminosity is well constrained by the spectral analysis, while the highest uncertainty is associated with $r$ and $n$. An upper limit for $r$ can be obtained writing $N_H\simeq n\Delta r$, with $\Delta r$ as the thickness of the absorber along the line of sight. Then, assuming that $\Delta r<r$ we can write $r\leq\frac{L_{\text{ion}}}{\xi N_H}$.\par
From the spectral analysis we see that the assumption of a single partial covering ionized medium is sufficient to give a good agreement with data. Therefore we can exclude that the absorption is due to a medium with a large range of ionization parameter and/or distributed along a broad range of distances from the center. If this were the case, the fit with a single absorbing component would not give the agreement that we obtain. Consequently, we can assume, as a first order approximation, that the absorber is composed by a single medium, whose thickness is much smaller than its distance from the center, namely $\Delta r<<r$, and whose ionization parameter is contained in a small range $\xi_0-\Delta\xi<\xi<\xi_0+\Delta\xi$, with $\Delta\xi\ll\xi_0$.
In this case we can write

\begin{equation}
r<< \dfrac{L_{\text{ion}}}{\xi N_H}
\end{equation}
In order to estimate $L_{\text{ion}}$, we assumed that the ionizing continuum between 13.6 eV and 13.6 keV follows a power law, with the same photon index found in the X-ray band. In XSPEC we used the task \texttt{clumin} to obtain
$$
L_{\text{ion}}=L_{\text{13.6 eV - 13.6 keV}}=(1.24^{+0.17}_{-0.15})\times 10^{45} \text{erg/s}
$$
which gives
$$
r<<r_{max}=185 \, \text{pc.}
$$
Alternatively, we can also parametrize $\Delta r=\alpha r$, with $\alpha$ an unknown constant. The photoionization model that we considered in our spectral analysis assumes a thin shell of absorbing material and therefore we can assume $\alpha <<1$. Thus we can write
\begin{equation}
\xi=\dfrac{L_{\text{ion}}}{nr^2}\simeq\dfrac{L_{\text{ion}}}{\frac{N_H}{\Delta r}r^2}=\alpha \dfrac{L_{\text{ion}}}{N_H r}
\end{equation}
and hence, for $N_H=10.1\times 10^{22}\text{cm}^{-2}$ and $\log{\xi}=1.33\,\text{erg}\,\text{s}^{-1}\,\text{cm}$
\begin{equation}
\label{r_abs}
r=\alpha \dfrac{L_{\text{ion}}}{\xi N_H} \simeq 150 \alpha \Big(\dfrac{L_{\text{ion}}}{10^{45}\text{erg/s}}\Big) \text{pc}
.\end{equation}
For comparison, we can consider the location of the BLR, which can be deduced using the H$\beta$ FWHM=4160 km/s=$v_{BLR}$ (\citealp{2002ApJ...571..733V}). Assuming a Keplerian orbit around the SMBH, we have
\begin{equation}
r_{BLR}\sim \dfrac{GM_{BH}}{v_{BLR}^2}\sim 0.2 \, \text{pc}.
\end{equation}
Therefore, for $\alpha \sim 10^{-3}$ the ionized partial covering absorber can be identified with the BLR, while for $10^{-3} \lesssim \alpha <<1 $ the absorber is more likely identified with the obscuring torus. Furthermore, we can use the definition of the ionization parameter $\xi$ to write the distance $r$ as a function of the density $n$ of the absorber, obtaining:
\begin{equation}
r=0.22\,\Big(\frac{n}{10^8\text{ cm}^{-3}}\Big)^{-1/2}\Big(\frac{L_{\text{ion}}}{10^{45}\text{erg/s}}\Big)^{1/2}\text{pc.}
\end{equation}
If we consider a density $n=10^9$ cm$^{-3}$, typical of the BLR clouds (e.g. \citealp{sned}), and using $L_{\text{ion}}=1.24\times 10^{45} \text{erg/s}$, the distance of  the absorber is $r\sim 0.1$  pc. We notice that this distance is comparable with the previous estimate of the location of the BLR.
\subsection{Covering factor and filling factor of the absorber}
Adopting the zxipcf model, we obtained a covering factor $f=0.88$, meaning that the absorber is not homogeneous, but rather has a clumpy structure. In the following, we show how we can relate the covering factor $f$ with the filling factor $\mathcal{F}$. Assuming that the absorber is composed by several distinct clouds , we can define $N_0$ the number of clouds per unit volume and $\bar{V}$ the average volume of a single cloud. Thus we can write
$$
\mathcal{F}=\dfrac{\text{volume occupied by the clouds}}{\text{total volume}}= N_0 \bar{V}
$$
The covering factor, instead, is defined as
$$
f=\dfrac{\Delta \Omega_{abs}}{\Delta \Omega}
$$
with $\Delta \Omega$ the solid angle subtended by the primary X-ray source, while $\Delta \Omega_{abs}$ is the portion of this solid angle covered by the absorber. If we call $\bar{L}$ the typical dimension of a single cloud, then $\bar{L}=\sqrt[3]{\bar{V}}$ and we can distinguish two cases: 1) the thickness of the absorber $\Delta r \sim \bar{L}$ and 2) $\Delta r >> \bar{L}$. In the first case, since the linear dimension of the clouds $\bar{L}$ is of the same order of magnitude of the thickness of the absorber, we can assume that the clouds do not overlap along the l.o.s.. Thus the filling factor can be rewritten as
\begin{equation}
\mathcal{F}=\dfrac{(\Delta \Omega_{abs}r^2 )\bar{L}}{(\Delta \Omega r^2)\Delta r}\sim f.
\end{equation}
We find an interesting consequence: given a thin absorbing shell with a certain covering factor, we have $\mathcal{F}\sim f$, independently on the distance from the primary source. This reasoning is no more valid for a thick shell, namely when $\Delta r >> \bar{L}$. We can consider, for instance, a thin shell with a certain filling factor $\mathcal{F}= N_0 \bar{V}\sim f$. Then, if we imagine a thick shell with the same value of $\mathcal{F}$, several clouds would overlap along the l.o.s. and  a greater fraction of $\Delta \Omega$ would be covered, implying $\mathcal{F}<f$. Thus we can conclude that, even if $\Delta r \sim \bar{L}$ or $\Delta r >> \bar{L}$, the covering factor $f$ represents always an upper limit for the filling factor $\mathcal{F}$. In our case we have
$$
\mathcal{F} \leq 0.88.
$$
We note that, even if in the thin shell limit we could assume $\mathcal{F}\sim f$, $N_0$ and $\bar{V}$ would still remain degenerate.
\subsection{Location and geometry of the reflector}
Furthermore, analyzing the Fe K$\alpha$ line, we can infer useful information about the location of the reflecting material. Firstly, we note that the best fit value of the line centroid is $6.41$ keV, which already indicates qualitatively that the bulk of the reflector is lowly ionized. In order to quantify the level of ionization we refer to \cite{1993MNRAS.262..179M}, who studied the physical properties of an accretion disk illuminated by an ionizing continuum. In particular, in Fig. 1 of \cite{1993MNRAS.262..179M} the centroid of the Fe K$\alpha$ line is plotted as a function of the ionization parameter $\xi$. Combining this figure and the results of our spectral analysis, we can affirm that, at $90\%$ level of confidence, $E_{ga}<6.5$ keV, which implies $\xi<400\,\text{erg}\,\text{s}^{-1}\,\text{cm}$. Thus, calling $\xi_R$, $n_R$ and $r_R$ the ionization parameter, the density and the distance of the reflector, respectively, we can write:
\begin{equation}
\label{xi_max}
\xi_R=\frac{L_{\text{ion}}}{n_R r_R^2}<\xi_{max}=400 \,\text{ erg}\text{ s}^{-1}\,\text{cm.}
\end{equation}
Analogously we can define $\xi_A$, $n_A$ and $r_A$ the ionization parameter, the density and the distance of the absorber, respectively. In general, the reflector and the absorber do not receive the same ionizing luminosity, due to the complexity of the nuclear geometry. In particular, if the reflector is much farther from the SMBH compared to the absorber, there is a time lag in the propagation of the ionizing continuum. Therefore we can parametrize  $L_{\text{ion,R}}=\lambda L_{\text{ion,A}}=\lambda n_A \xi_A r_A^2$, where the subscript “A” refers to the absorber, while “R” to the reflector. Using Eq. \ref{xi_max}, we can write
\begin{equation}
r_R>r_A\sqrt{\lambda \frac{n_A}{n_R} \frac{\xi_A}{\xi_{max}}}
\end{equation}
which can be rewritten using the Eq. \ref{r_abs} as
\begin{equation}
r_R>35 \alpha \sqrt{\lambda \frac{n_A}{n_R}} \frac{L_{\text{ion,A}}}{10^{45} \text{erg s}^{-1}}  r_A.
\end{equation}
\par
Such relation imposes a lower limit for the location of the reflector, but it contains unknowns that cannot be directly derived from our analysis, such as $\lambda$, $n_A$ and $n_R$. We can qualitatively deduce that the density of the reflector is likely much higher than that of the absorber, since the reflection component is not strong and the reflecting gas is less ionized than the absorbing gas. Nevertheless, this is not sufficient to derive a numerical constraint for a lower limit of $r_R$.\par
Another interesting information about the location of the reflector comes from the width of the Fe K$\alpha$ line, which appears not resolved. Indeed, if we let free the parameters $E_{ga}$ and $\sigma_{ga}$ during the fit, we find $\sigma_{ga}<120$ eV . This result is in accordance with what has been found in the literature for other BLRGs, which rarely show a relativistic broadened iron line. If we suppose that the broadening is due mostly to a Doppler kinematic effect, we can infer another constraint on the location of the reflector. Let us suppose that the reflector follows almost circular Keplerian orbits. In this configuration, part of the reflector moves toward the observer, part moves away, causing the Doppler broadening of the line. Since the orbital velocity increases at lower distances, we may say that an upper limit of $\sigma_{ga}$ implies a lower limit $r_{in}$ for the distance of the reflector. In order to find $r_{in}$, we make the rough approximation 
\begin{equation}
\frac{\Delta v_{//} }{c}\simeq \frac{\Delta E_{FWHM}}{E_{ga}}=\frac{2.35 \, \sigma_{max}}{E_{ga}}
\end{equation}
where $\sigma_{max}=120$ eV, $\Delta v_{//}=\Delta v \sin{i}$ and $\Delta v=2\, v_{orb}\Big|_{r_{in}}$. Therefore
$$
v_{orb}\Big|_{r_{in}}\sim \frac{c}{2\sin{i}}\frac{2.35 \, \sigma_{max}}{E_{ga}}= \beta \, c
$$
For $50\degree<i<70\degree$, we have that $0.024<\beta<0.030$.
Then, knowing that 
$$
\dfrac{v_{orb}(r_{in})}{c}=\sqrt{\dfrac{r_s}{2r_{in}}}=\beta
$$
with $r_s=\dfrac{2GM_{BH}}{c^2}$, we obtain
\begin{equation}
r_R>r_{in}=\frac{r_s}{2\beta^2}
\end{equation}
where $600 \, r_s \lesssim r_{in}\lesssim 900 \, r_s$.
Although, we stress that this lower limit for $r_R$ is related to the reflecting material that produces the observed narrow Fe K$\alpha$ line. In principle, a fainter and broader iron line can be present in the spectrum, but not detectable with the S/N of our data. It could be also possible that the inner part of the disk ($r_R\lesssim600r_s$) does not contribute considerably to the emission of Fe fluorescence because its level of ionization is too high and most of the outer electrons are ripped off the atoms.\par
The \texttt{pexmon} model does not modify sensibly the best fit values of $N_H$, $\log{\xi}$, $f$ and $\Gamma$. The reflection parameter R is constrained in the range $0.64<R<1.85$, having assumed an inclination angle in the range $50\degree<i<70\degree$. Such result is compatible with a scenario where the reflector covers most of the solid angle illuminated by the primary source. If also the dusty torus contributes to the reflection, a value of $R\sim1$ implies that the torus has a flattened geometry, meaning that almost all the radiation emitted by the primary source is reflected toward the observer. The scenario of a flattened torus is also supported by \cite{2009ApJ...705..298M}, who found an opening angle of only 15$\degree$. Anyway, the data quality is not good enough to infer further information about the geometry of the reflector.

\subsection{Variability of the reflection component}
The timing analysis of the source light curve does not show any significant evidence of variability of the total flux during the entire exposure. Then we investigated the variability of the hardness ratio.
Two intervals, that we called B and C, of the duration of 6 ks and 4 ks respectively, show an HR significantly higher than the average value. We analyzed the spectrum in these intervals, in order to test the variability of the individual spectral components. \par
Neither the fit parameters associated with the absorber nor the properties of the power law change significantly when the HR is higher. 
Instead, the reflection component shows evidence of variability, especially in the interval B. If the Fe K$\alpha$ line is modeled with a Gaussian, its normalization increases at least of a factor of two in B. As a consistency check, also using the \texttt{pexmon} model the reflection factor has an average value of $\expval{R}\lesssim 1$ in A, while $R>1$ in B.\par
In order to explain the variability of the reflection component, we notice that the intensity of this last is directly proportional to the incoming primary flux. In the case of the Fe K$\alpha$ line, since the fluorescence is due to the absorption of X-ray photons with E$\geq$7 keV, a variability of the incoming flux above 7 keV ($F_{>7\text{ keV}}$) causes the variability of the line intensity. Therefore, we propose two different scenarios: i) We interpret the reflection variability as an effect of Compton thick clouds rapidly moving in the immediate vicinity of the X-ray primary source. If such clouds intercept the flux directed to the reflector, the flux $F_{>7\text{ keV}}$ is partly suppressed, otherwise it increases. In such scenario the variability timescales of the reflection component are related to the dynamical time scales of the obscuring clouds. Alternatively we could suppose that the obscuring clouds are not located between the primary source and the reflector, but rather between the reflector and the observer. Though, in this case, two drawbacks arise. First, if the obscuring clouds are on the line of sight, we should observe at the same time a variability of both the reflection component and the column density $N_H$, which is not our case. Second, since we have derived a lower limit for the distance of the material that produces the Fe K$\alpha$ line, the dynamical time scales of the obscuring clouds would be too long to explain the variability of the reflection on time scales of $\sim10^4$ s. ii) The reflection variability is directly caused by an intrinsic variability of the primary source luminosity, but since the narrow Fe K$\alpha$ line is produced at a distance $R_R\gtrsim 600\, r_s$, there is a net time lag between the arrival time of the primary continuum and the reflected component. The time lag would be at least of 60 days, which could explain why in a single exposure of $\sim$ 17 hrs, we do not observe an increase of the primary flux followed by the increase of the reflection component.

\section{Conclusions}
Through the X-ray analysis of an observation performed with the XMM-Newton observatory, we have obtained novel information about the geometry, kinematics, and physical state of the regions surrounding the accreting SMBH at the center of the BLRG \object{PKS 2251+11}. The X-ray spectrum is consistent with a primary component described by a power law, which is absorbed by an ionized partial covering medium. Considering a density of the absorber typical of the BLR, its distance from the SMBH is of the order of  $R=0.1$ pc. This estimate is comparable with the location of the BLR, estimated through the virial theorem and knowing the FWHM of the H$\beta$. Alternatively, if we assume lower values of density, it is not excluded that the absorber is located farther from the SMBH, in the zone of the dusty torus. The absorber is likely clumpy, in other words, composed by distinct clouds with a covering factor $f\simeq 90\%$ and a filling factor $\mathcal{F}\lesssim f$. The spectral analysis confirms the presence of an unresolved Fe K$\alpha$ line and the bulk of the emission occurs at a distance $r_R>600r_s$. \par
Besides the intensity of the Fe K$\alpha$ line, no other spectral parameters show significant variability during the observation. Due to the considerable distance between the reflector and the SMBH, the variability of the Fe K$\alpha$ intensity is likely due to past variations of the primary flux. Nonetheless, a rearrangement of the absorber geometry may also be a valid alternative.\par
An interesting follow up investigation could be an observation with the NuSTAR observatory of the high-energy part ($E>10$ keV) of X-ray spectrum. With such analysis we could see if the power law has the same photon index also at higher energies, understanding in this way whether other components, such as the radio jet, can contribute to the X-ray emission of this AGN. Furthermore the energy of the power law cut-off (typically around 100 keV) is directly connected with the temperature of the electrons of the hot corona.\par
It would be also interesting to extend this work to a multi-epoch X-ray observation of \object{PKS 2251+11}, exploring the variability on timescales of days or months. For instance the scenario of a delayed response of the reflection component after a variation of the primary flux could be further tested. The light curves of the primary continuum and the reflection component could be cross-correlated, similarly to the method adopted in reverberation mapping of the BLR in the optical/UV band. Possible evidence of delay could further constrain the location of the reflector. With an X-ray multi-epoch study we could also test the way in which the geometry of the absorber influences the variability of the spectrum at energies $E<2$ keV.

\begin{acknowledgements}
      S. Ronchini, who carried out this work during his master thesis at the University of Rome "Tor Vergata", thanks the other authors for the valuable support.
      F. Tombesi acknowledges support by the Programma per Giovani Ricercatori - anno 2014 “Rita Levi Montalcini”. G.Bruni acknowledges financial support under the INTEGRAL ASI-INAF agreement 2013-025. R01. The authors thank the anonymous referee for the constructive comments.
\end{acknowledgements}

\bibliographystyle{aa}
\bibliography{biblio_new}

\end{document}